\documentclass[]{aa}
\usepackage{graphics,latexsym,psfig}
\usepackage{graphicx}

\def\farcs{\hbox{$.\!\!^{\prime\prime}$}}  % Fractions of arcseconds
\def\asec{\ifmmode ^{\prime\prime}\else$^{\prime\prime}$\fi}
\def\amin{\ifmmode ^{\prime}\else$^{\prime}$\fi}
\def\degs{\ifmmode ^{\circ}\else$^{\circ}$\fi}
\def\etal{{et\,al. }}

\begin{document}

\title{Constraining the GRB Collimation with a Survey for Orphan Afterglows}

\titlerunning{Constraining the GRB Collimation}

\author{
A. Rau \inst{1}\inst{2}        \and
J. Greiner \inst{2} \and
R. Schwarz \inst{3}
}
\offprints{A. Rau, arne@astro.caltech.edu}

\institute{Division of Physics, Mathematics and Astronomy,  California Institute of Technology, Pasadena, CA 91125, USA \and Max-Planck-Institut      f\"ur      extraterrestrische      Physik,
  Giessenbachstrasse,   85748  Garching,   Germany   \and  Astrophysikalisches
  Institut Potsdam, An der Sternwarte 16, 14482 Potsdam, Germany
}

% --------------------------------------------------------------------------
\date{Received Oct 7 2005 / Accepted Nov 22 2005}

\abstract{Gamma-ray   bursts   are   believed   to  be   produced   in
highly-relativistic collimated outflows.  Support for this comes among
others from  the association  of the times  of detected breaks  in the
decay  of afterglow  light curves  with the  collimation angle  of the
jets.  An alternative approach to  estimate a limit on the collimation
angle   uses   GRB   afterglows   without   detected   prompt-emission
counterparts.  Here  we report on  the analysis of a  dedicated survey
for the search of these  orphan afterglows using the Wide Field Imager
at  the 2.2\,m  MPI/ESO telescope  at  La Silla,  Chile. We  monitored
$\sim$12\,deg$^2$ in  up to 25 nights  typically spaced by  one to two
nights with  a limiting magnitude  of $R$=23. Four  previously unknown
optical transients were discovered  and three of these associated with
a flare  star, a  cataclysmic variable and  a dwarf nova.   The fourth
source shows  indications for an  extragalactic origin but  the sparse
sampling  of the light  curve prevents  a reliable  classification. We
discuss the results in the context of the collimation of GRBs.
} \maketitle

% --------------------------------------------------------------------------
\section{Introduction} 

There is  now conspicuous observational and  theoretical evidence that
the radiation of cosmic gamma-ray  bursts (GRBs) is produced in highly
relativistic  collimated  outflows.  A  jet  geometry  was  originally
invoked as a solution for  the ``energy crisis'' by reducing the total
energy output  of a GRB  by a factor of  $\Omega_\gamma$/4$\pi$, where
$\Omega_\gamma$ is  the solid angle into which  gamma-rays are emitted
(e.g.,  Rhoads   1997;  Fruchter   \etal  1999).   Evidence   for  the
collimation was  provided by  the theoretically predicted  (e.g., Sari
1999;  Gruzinow  1999;  Ghisellini   \&  Lazzati  1999)  and  observed
polarization evolution of optical afterglows (e.g., Covino \etal 1999;
Wijers \etal 1999; Greiner \etal 2003).

Prominent signatures  of the jet  geometry were identified already  earlier in
the broad-band breaks observed in the optical and radio afterglow light curves
of several long duration GRBs  (e.g., Stanek \etal 1999; Harrison \etal 1999).
The opening angles,  which have been inferred from these  ``jet breaks'' for a
number of  afterglows, vary from 1\degs\  to more than 25\degs,  with a strong
concentration near 4\degs\ (Frail \etal 2001).

%However,  breaks obtained  from the
%radio  afterglow light curves  occur typically  much later  than those
%observed  at  optical  wavelengths.   This suggests  that  either  the
%optical transients  typically radiate into a smaller  solid angle than
%the radio transients or that  the radio breaks are not associated with
%the jet opening angle.

A consequence of the collimation is that the prompt $\gamma$-ray emission will
be detected if the  viewing angle of the observer is equal  to or smaller than
the opening  angle of the jet,  $\theta_{jet}$.  This implies that  only for a
fraction of  all bursts  in the Universe  $\gamma$-ray photons will  reach the
Earth. The total GRB rate will be higher than the observed rate by a factor of
roughly  $\theta_{jet}^{-2}\propto\Gamma^2$   in  case  of   a  universal  jet
structure, where  $\Gamma$ is the  bulk Lorentz factor  of the ejecta.   For a
quasi-universal    Gaussian-type   jet,   the    rate   will    be   generally
smaller. Nevertheless, GRBs which are not pointed directly at the observer can
in  principle  be  discovered  through  their afterglow  radiation  at  longer
wavelengths.

In the standard internal-external fireball model (e.g.  Rees \& Meszaros 1992,
1994) the afterglow is produced  when the initially highly relativistic ejecta
plows  into the ambient  medium.  The  afterglow emission  is radiated  into a
solid angle of $\Omega_A$$\sim$1/$\Gamma$, along the line of motion.  When the
jet  decelerates, $\Omega_A$  increases until  it reaches  4$\pi$.  Therefore,
so-called ``off-axis'' orphan afterglows can  be detected for bursts which are
beamed outside of the field of  view of the observer (e.g., Rhoads 1997; Perna
\& Loeb  1998; Dalal \etal  2002). The light  curves of these  off-axis orphan
afterglows will initally be faint, brightening up to a viewing angle dependent
maximum and become similar to regular GRB afterglows after the jet break later
on (Rhoads 1999; Nakar \& Piran 2003).

It is theoretically  reasonable to assume that the  optical afterglow might be
emitted  from a slower  moving material  (a lower  $\Gamma$) than  the initial
$\gamma$-ray  beam.  Therefore,  so-called ``on-axis''  orphan  afterglows are
expected when the narrow $\gamma$-ray  emission misses the observer by a small
amount but the wider optical emission region falls within the observation cone
(Nakar \& Piran 2003).  These  afterglows will exhibit similar light curves as
regular afterglows with detected prompt emission.

Observations of orphan afterglows can  help to study the initial opening angle
of  the jets  and to  place a  constraint on  the collimation  of  the optical
afterglow emission (Rhoads 1997).   Especially on-axis orphans are suitable as
they  are substantially  brighter than  off-axis  orphans and  thus easier  to
detect in a dedicated survey (Nakar \& Piran 2003).  Additionally, Dalal \etal
(2002) pointed out, assuming  an uniform jet with constant  jet-break time and
luminosity at  the break time for  an on-axis observer, that  for small angles
the  afterglow effective  beaming angle  scales  with the  jet opening  angle.
Therefore,  the  number of  detectable  off-axis  orphan  afterglows would  be
independent  of the jet  opening angle  and similar  for moderately  wide jets
($\sim$20\,\degs)  and  for  arbitrarily  narrow  jets  ($<$0.01\,\degs).   In
contrast,  Totani \&  Panaitescu (2002)  predict  a strong  dependence of  the
orphan rate on  the jet opening angle assuming a constant  total energy in the
afterglow jet.

\bigskip 

A small number  of surveys dedicated to the search  of untriggered optical GRB
counterparts were performed over the past years.  No candidate event was found
in 125\,hrs monitoring of a field of 256\,deg$^{2}$ with ROTSE-I to a limiting
magnitude of 15.7  (Kehoe \etal 2002).  Vanden Berk  \etal (2002) searched for
color-selected  transients within  1500\,deg$^{2}$  of the  Sloan Digital  Sky
Survey (SDSS)  down to  R=19 and  found only one  unusual transient  which was
later identified  as a radio-loud  AGN exhibiting strong  variability (Gal-Yam
\etal 2002).  The automated RAPTOR  wide-field sky monitoring system allows to
image 1300\,deg$^{2}$  at a time down  to a magnitude  of $\sim$12.5 (Vestrand
\etal 2004).   A couple  of interesting optical  transients were found  in the
$B$, $V$  and $R$-band Deep  Lens Survey transient  search, within an  area of
0.01\,deg$^{2}$ yr  with a limiting magnitude  of 24.  None of  those could be
positively  associated with a  GRB afterglow  (Becker \etal  2004).  Recently,
Rykoff \etal  (2005) performed  a search using  the ROTSE-III  telescope array
without detecting any candidate afterglow  events.  They placed an upper limit
on the  rate of fading  optical transients with quiescent  counterparts dimmer
than $\sim$20th magnitude of less than 1.9\,deg$^{-2}$ yr$^{-1}$.

In this  paper we present the  results of a survey  for optical orphan
afterglows  with a  wide-field  imaging instrument  and  will give  an
estimate of  the GRB  collimation derived from  the detection  rate of
on-axis orphan  afterglows.  The paper  is structured as  follows.  In
Sect.~2  we describe the  observational strategy,  instrumentation and
selected survey  fields as  well as the  data reduction  and transient
detection method.   The candidate transients are  presented in Sect.~3
and the transient detection efficiency in Sect.~4. A discussion of the
results in the context of GRB collimation is given in Sect.~5.

% --------------------------------------------------------------------------
\section{Observations and Data Reduction}

\subsection{Strategy, Instrumentation \& Survey Fields}

A search  for GRB afterglows or  other transient phenomena  in the sky
requires a thorough strategy due to the random occurence of the events
in space and time.  As a continuous monitoring of a large field with a
large  aperture telescope  could only  be considered  overambitious in
many  cases, we  developed our  strategy  using the  knowledge of  the
properties of observed optical afterglows  at the time when the survey
was proposed (1999).

The  primary idea  was  to take  multiple  deep observations  of  a number  of
selected  sky fields. The  observing scheme  was chosen  such that  the survey
would be sensitive enough ($R$$\sim$23\,mag) to provide the detection of a GRB
orphan afterglow in at least two  epochs together with earlier and later upper
limits. To  combine the  availability of a  large aperture  telescope together
with the  observed brightness decay of  GRB afterglows, we  decided to perform
consecutive observations of a given field  in every 2nd night.  Over this time
span,  on-axis  orphan  afterglows  will   in  most  cases  be  brighter  than
$R$=21. Instead, off-axis orphans are expected to be fainter and will start to
dominate below that magnitude (Nakar \& Piran 2003).

We  obtained imaging data  during three  periods (May-June,  August \&
September-October) in 1999 and monitored  7 different sky fields in up
to  25  nights each.   Each  field  is composed  of  2  to 8  separate
sub-fields  and  a  total   of  38  sub-fields  were  selected.   This
corresponds  to  an  area  of  $\sim$12 square  degrees  being  imaged
throughout the campaign. A list of all fields together with the number
of sub-fields and  the maximum number of observing  nights is provided
in Table~\ref{tab:obslog}.  Note that  due to changes of the observing
conditions during a night not  always a complete monitoring of a given
field could be accomplished.

A total of 39\,nights were scheduled using the Wide Field Imager (WFI)
at the  MPI/ESO 2.2\,m telescope in  La Silla, Chile.   Due to weather
constraints  only  31  nights  could  at  least  partly  be  used  for
observations.   Unfortunately,  the  lost  nights  caused  unfavorable
interruptions  of the  otherwise gap-less  observing schedule  of each
period and lead to a significantly decreased detection sensitivity for
orphan afterglows.

\begin{figure}[h]
\includegraphics[width=0.69\linewidth,angle=270]{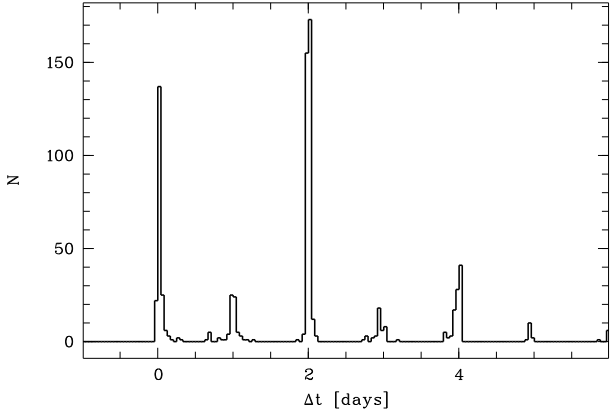}     
\caption{The distribution of time between consecutive observations of
  individual sub-fields with 1\,hr binning.}
\label{fig:tDifObs}
  \vspace{-0.5cm}
\end{figure}

The   distribution  of   the  time   delay  between   two  consecutive
observations     of    a     given    sub-field     is     shown    in
Figure~\ref{fig:tDifObs}. While  the majority  of the survey  could be
observed  with  the  proposed  gap  of two  days  between  consecutive
observations,   $\sim$15\,\%  suffered   from  larger   gaps   due  to
inappropriate  weather conditions.  In addition  to the  regular 2-day
schedule  a small  number of  nights were  included in  which multiple
observations of selected  fields (F1, F3 \& F5)  were performed within
one  night.  This  would allow  to identify  and  distinguish possible
short-term variable sources (e.g. CVs) in these fields.

% --------------------------------------------------------------------------
% --------------------------------------------------------------------------

\begin{figure}[h]
\includegraphics[width=0.69\linewidth,angle=270]{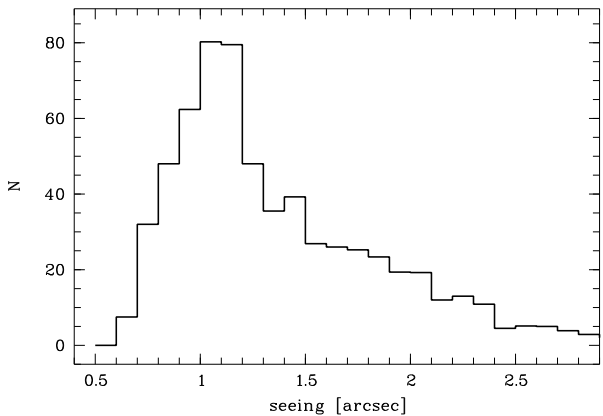}     
\caption{Seeing distribution for all pointings. }
\label{fig:seeing}
  \vspace{-0.5cm}
\end{figure}

The photometry was taken with WFI mainly in the $R$-band together with a small
number of pointings performed using the $V$ and $I$-band filters. The typical
exposure time was 420\,s per pointing.  The instrument consists of a mosaic of
4$\times$2 CCDs each  2046$\times$4128 pixel in size which,  together with the
plate scale of  0.238 arcsec/pixel, provided a sky  coverage of $34'\times33'$
per image.   A gap of  $\sim$2 arcsec exists  between neighboring CCDs  of the
imager.   Throughout  the data  acquisition  the  observing conditions  varied
strongly (see Figure~\ref{fig:seeing}). The majority of the data were taken at
a seeing of  $\sim$1'' but a small fraction of  the observations suffered from
significantly worse seeing. Due to the  low limiting magnitude and the loss of
all but the brightest sources  we neglected date taken under seeing conditions
above 2\farcs4 ($\sim$10  pixel FWHM). This corresponds to  an additional loss
of $\sim$10\,\% of the campaign data.

% --------------------------------------------------------------------------
% --------------------------------------------------------------------------

 The selection of  the fields was more or less  random with respect to
the search for  orphan afterglows with the exception  of the field F6.
This field is centered on Selected  Areas 113 and includes a number of
Landoldt   standard  stars   which  allow   an   absolute  photometric
calibration of  the survey (Landoldt  1992; see below).   F6, together
with  a  second   field  (F4),  is  also  covered   by  the  SDSS\footnote{
http://www.sdss.org/}.   This  provided  us  with the  opportunity  to
identify  possible   transient  sources   in  these  fields   down  to
$R$$\sim$22\,mag  using the  SDSS multi-band  informations. Particular
care was taken to avoid stars brighter then $\sim$12th magnitude which
would not  only saturate the  detector but also leave  effected pixels
less  sensitive  for  subsequent  images. Furthermore,  high  ecliptic
lattitudes were favored  to maximize the Moon distance,  and thus make
the fields  observable over a  large fraction of an  observing period.
Finally,  significant  foreground extinction  was  avoided during  the
field selection.

% --------------------------------------------------------------------------
\begin{table}
  \caption{Observation  log.  The first  four columns  provide the  ID, central
  coordinates and number of sub-fields.  The last column represents the amount
  of nights with seeing $<$2\farcs4  in which pointings of a given field
  were obtained.}
  \begin{tabular}{ccccc}
    \hline\hline
    Field & RA(2000) & Dec(2000) & sub-fields & \# of nights \\
    \hline
    F1  & 01h32m & --43\degs12\amin & 4 & 15 \\
    F2 & 03h33m & --27\degs37\amin & 4 & 12 \\
    F3 & 13h28m & --21\degs40\amin & 8 & 11 \\
%    4 & 14 29 42.9 & --62 40 46 & $68'\times33'$ & 14 \\
    F4 & 16h20m & +04\degs00\amin & 8 & 12 \\
%    6 & 18 49 07.5 & --03 20 30 & $68'\times99'$& 23 \\
%    7 & 19 25 31.8 & +02 47 40 & $68'\times33'$ & 21 \\
    F5 & 21h26m & --43\degs22\amin & 8 & 23 \\
    F6 & 21h41m & +00\degs30\amin & 2 & 25 \\
    F7 & 21h52m & --27\degs32\amin & 4 & 21 \\
    \hline\hline
  \end{tabular}
  \label{tab:obslog}
  \vspace{-0.5cm}
\end{table}
% --------------------------------------------------------------------------

\subsection{Data Reduction}

Nearly 700 images were obtained throughout the survey comprising $\sim$130\,GB
of raw data.  For the automatic reduction and analysis of this large amount of
data a  Perl-based pipeline was developed  which uses a number  of well tested
astronomical software packages.  The basic image reduction was performed using
{\it  IRAF}\footnote{{\it  IRAF}  is   distributed  by  the  National  Optical
Astronomy Observatories, which are operated by the Association of Universities
for Research in Astronomy, Inc., under cooperative agreement with the National
Science Foundation.}{\it  /MSCRED}.  For each  night of observations  a common
bias frame  was produced and subtracted  from the science  images.  Flat field
correction was performed using daily super-sky-flats produced from all science
observations of a given night  without significant illumination by the Moon or
other bright sources.

\begin{figure}[h]
\includegraphics[width=0.73\linewidth,angle=270]{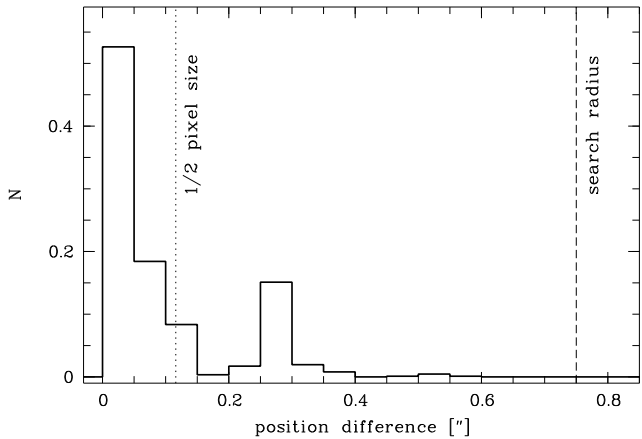}     

\caption{Normalized distribution of the distances of $\sim$40 isolated sources
  detected  during  the  whole  survey  in  the field  F6-1  compared  to  the
  respective reference positions obtained from an observation taken on May 31,
  1999 (solid  line). The  dotted and  dashed lines indicate  the size  of 1/2
  detector  pixel and  the search  radius  for the  source tracing  algorithm,
  respectively.  The selected sources  span $\sim$5\,mag in brightness and are
  distributed over the entire detector field of view.}
\label{fig:aA}
  \vspace{-0.5cm}
\end{figure}

The  astrometric   solutions  were  obtained   separately for each of the 8 CCDs  with   the  {\it
WIFIX/ASTROMETRIX}\footnote{http://www.na.astro.it/$\sim$radovich/wifix.htm}
package by comparing the positions  of detected sources with those compiled in
the USNO-A2.0 catalog (Monet \etal 1998).  The resulting astrometric precision
is indicated  in Figure~\ref{fig:aA} where  the difference in the  position of
isolated sources  detected in  a sub-field of  F6 in multiple  observations is
shown.  About 80\,\% of the sources lie within 1/2 detector pixel (0\farcs119)
compared to a  reference observation taken on May 31,  1999.  More than 99\,\%
of the sources are detected within a circle of 0\farcs4 radius.  Note that the
double peaked shape of  the distribution in Figure~\ref{fig:aA} is artificial,
resulting from the  numerical rounding of the pixel  coordinates in the source
detection algorithm.

% --------------------------------------------------------------------------
% --------------------------------------------------------------------------

The source detection in the images  was performed in each of the 8 WFI
CCDs separately.   We used the  {\it IRAF/DAOPHOT} package  to measure
the source  flux inside a  Gaussian-shaped point-spread-function.  The
search for transient sources does not gain significant benefit from an
absolut  photometric  calibration which  would  be  connected with  an
unavoidable additional systematic zero-point uncertainty.  Instead, we
obtained   the  variability   information  using   the   technique  of
differential  photometry.  We  selected  an ensemble  of  at least  20
local, non-saturated, non-variable reference stars for each sub-field.
By deriving the  median brightness offset of these  stars with respect
to the brightness of the same stars obtained in a reference image of a
given  sub-field (typically  the one  with the  smallest  seeing), the
photometric offset  between the  two observations could  be estimated.
This was successively  done for all observations of  a given sub-field
and thus provided a common photometric zero-point for all pointings of
the respective sub-field.

Absolut photometric calibration was  obtained based on observations in
one  photometric night  (October 12,  1999)  of the  field SA113  (our
F6). This field contains a number of Landoldt standard stars with well
tabulated  optical photometry  (Landoldt  1992)\footnote{The tabulated
Cousin $R$-band magnitudes were transformed to the WFI $R$-band filter
system   using   the   colour    and   extinction   terms   given   at
\footnote{http://www.ls.eso.org/lasilla/Telescopes/2p2T/E2p2M/WFI/zeropoints/}.}.
As all photometric standard stars were saturated in the regular 420\,s
image taken  for our  survey, we obtained  a shorter  exposure (20\,s)
during the same night.  In this 20\,s frame six non-saturated standard
stars were contained and the photometric zero point for this image was
obtained (systematic uncertainty of $\Delta R$=0.1\,mag). This allowed
to produce a sample of secondary standard stars in the field down to a
limiting  magnitude  of   $R$$\sim$21  corresponding  to  a  5$\sigma$
detection.

The field F6 was observed in nearly  all nights of the survey. This allowed to
obtain accurate photometric calibration for  the other six fields based on the
zero points derived from the secondary standards in F6 in a common night. For
each field one  photometric night was chosen and  the photometric calibration
of all  further observations were calibrated  with respect to  this night. The
photometry in  all fields was  additionally corrected for  Galactic foreground
extinction ($A_R$=0.02--0.25) (Schlegel \etal 1998).

\begin{figure}[h]
\includegraphics[width=0.73\linewidth,angle=270]{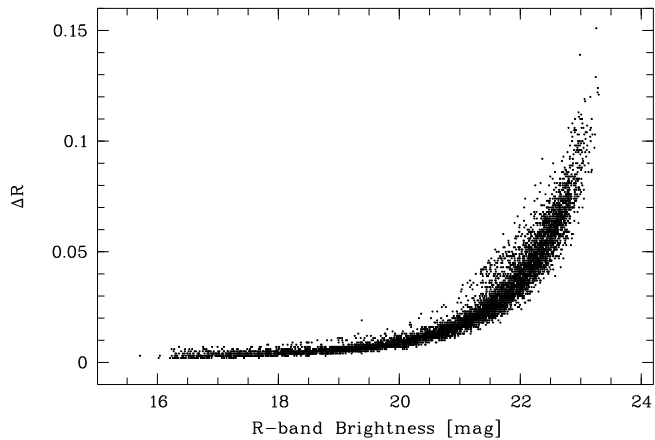}     
\caption{Photometry obtained for an observation of the field F4 on May
22, 1999 with each dot  representing one source. The seeing during the
time  of  the  imaging  was  0\farcs8  and  a  limiting  magnitude  of
$R$$\sim$23\,mag with  10$\sigma$ was reached.   Sources brighter than
$R$$\sim$16\,mag are saturated.}
\label{fig:photQual}
  \vspace{-0.5cm}
\end{figure}

Figure~\ref{fig:photQual}  shows  the  resulting  photometric quality  for  an
example  field (F4)  obtained under  good seeing  conditions  (0\farcs8).  The
uncertainties  represent  the  detection  uncertainties  in  the  instrumental
magnitude  system and  do not  include  the uncertainties  resulting from  the
absolute  photometric  calibration  ($\Delta  R$$\sim$0.1\,mag).   A  limiting
magnitude of  $R$$\sim$23 at  10$\sigma$ was reached  in this  observation. As
discussed  above,  the observing  conditions  varied  strongly throughout  the
survey.  Thus, the achieved  limiting magnitude ranged between $R$$\sim$19 and
$R$$\sim$23, depending on the  seeing and Moon illumination.  Sources brighter
than $R$$\sim$16 were saturated in a good night.

% --------------------------------------------------------------------------
% --------------------------------------------------------------------------

\subsection{Transient Detection Method}

For each  of the  seven fields  a master table  including the  coordinates and
magnitudes of all  detected sources in all observations  was produced. Herein,
we assigned detections within a  distance of 0\farcs75 to a detection obtained
in a reference image together and consider this as one source.

Light curves  were obtained for  all detected sources and  candidate transient
objects were selected  based on the deviation of their  light curve from their
mean   light  curve.    For   all  candidates   with   single  detections   or
$\Delta$$R$$>$0.75\,mag ($\sim$12000), light curve  plots as well as thumbnail
images of all pointings were produced.  These were examined by eye in order to
remove spurious transients arising from nearby bright stars, extended objects,
stray light effects at the edge of the FoV, bad focus or detections which were
consistent  with faint  stars  at  the limiting  magnitude  of the  individual
pointings. The  strategy of the  survey was aimed  to catch a  possible orphan
afterglow in at least  two consecutive observations.  Therefore, we considered
only sources  with at  minimum two detections  for further  investigations.  A
number of fast moving objects (e.g.  air planes, terrestrial satellites) could
be identified  by the trace  in the images  which they left during  the 420\,s
exposure.  Solar  system objects (e.g.  asteroids) are  especially abundant in
the survey fields  located close to the ecliptic plane (F3,  F6 \& F7).  These
sources  appear  either as  one-time  detections  or  show significant  motion
between subsequent  observations and could be identified  accordingly.  Of the
$\sim$12000 candidates  only four  remained and were  double checked  with the
positions       of       known       sources       compiled       in       the
SIMBAD\footnote{http://simbad.u-strasbg.fr/},
NED\footnote{http://nedwww.ipac.caltech.edu/}                               and
NSV\footnote{http://heasarc.gsfc.nasa.gov/W3Browse/all/gcvsnsvars.html}
databases  and  if possible  correlated  with the  SDSS  4th  release and  the
DSS\footnote{http://archive.stsci.edu/cgi-bin/dss\_form}.

% --------------------------------------------------------------------------
\section{Results}

\subsection{Candidate Transients}

Throughout the survey four new transient sources with detections in at
least two images were found. Based on the shape of the light curve and
amplitude we identified one  candidate cataclysmic variable, one dwarf
nova,  one  flare  star  and  one  candidate  extragalactic  transient
superimposed on an  underlying faint object. For the  latter an orphan
afterglow  nature is  considered  plausible.  Below  we  give a  brief
description  of  each  of  these  four  transient  sources  (see  also
Table~\ref{tab:candidates}).

% --------------------------------------------------------------------------
\begin{table}[h]
  \caption{Candidate transient sources. The minimum and maximum obtained
  brightnesses as well as the putative classification are given.}
  \begin{tabular}{lccl}
    \hline\hline
     \multicolumn{1}{c}{Name} & \multicolumn{2}{c}{Brightness [mag]} & \multicolumn{1}{c}{Putative} \\
     & min  & max & \multicolumn{1}{c}{classification}\\
     \hline
     J132653.8-212702 & 20.3$\pm$0.1 & 19.5$\pm$0.1 & CV, eclip. binary \\
     J132813.7-214237 & 21.3$\pm$0.1 & 19.9$\pm$0.1 & extragalactic \\
     J161953.3+031909 & 19.9$\pm$0.1 & 17.5$\pm$0.1 & dwarf nova\\
     J215406.6-274226 & 22.5$\pm$0.2 & 20.0$\pm$0.1 & flare star\\ 
   \hline\hline
  \end{tabular}
  \label{tab:candidates}
  \vspace{-0.5cm}
\end{table}
% --------------------------------------------------------------------------

\subsubsection{J132653.8-212702}

\begin{figure*}
\includegraphics[width=0.365\linewidth,angle=-90]{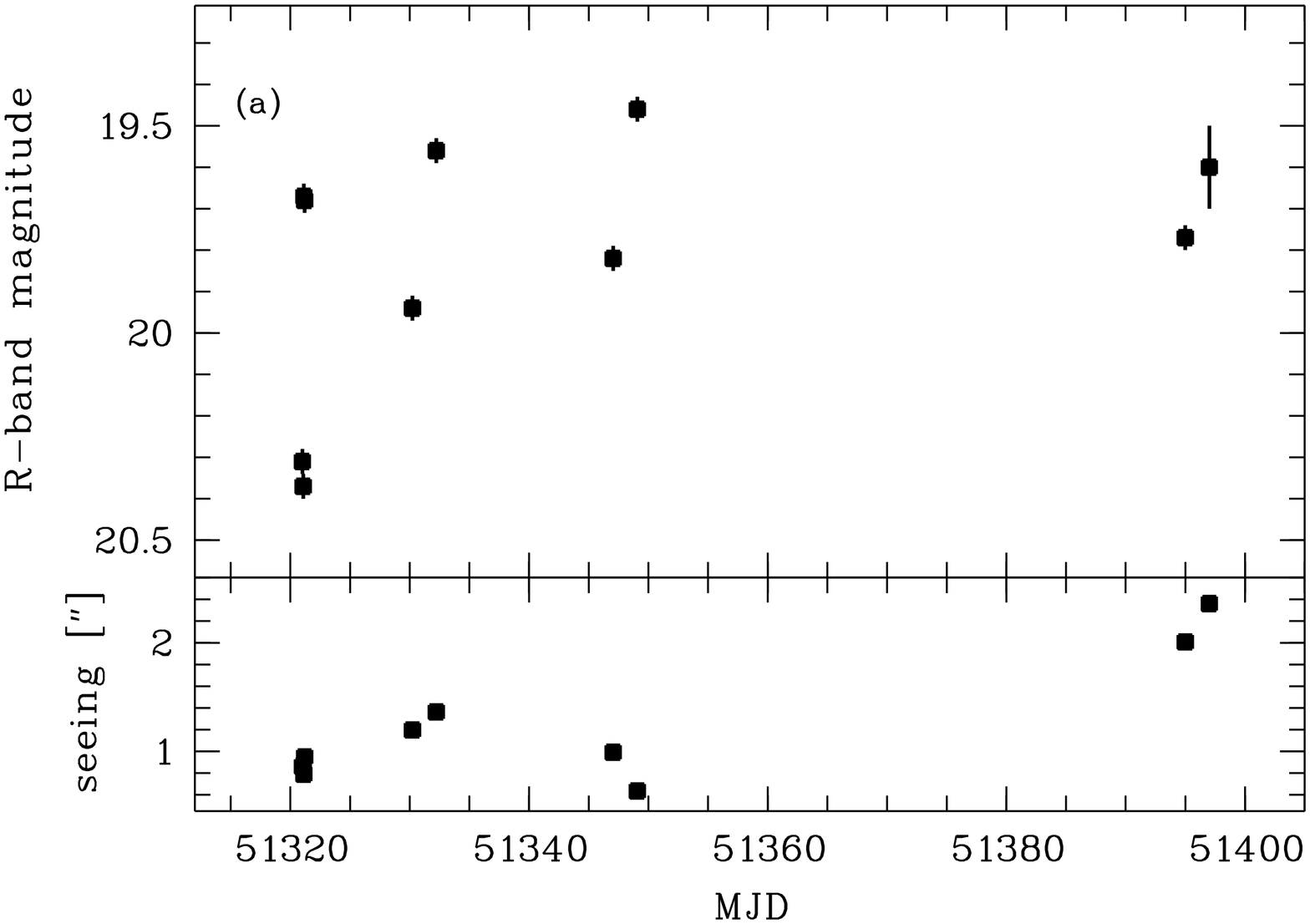}     
\hspace{-0.65cm}
\includegraphics[width=0.365\linewidth,angle=-90]{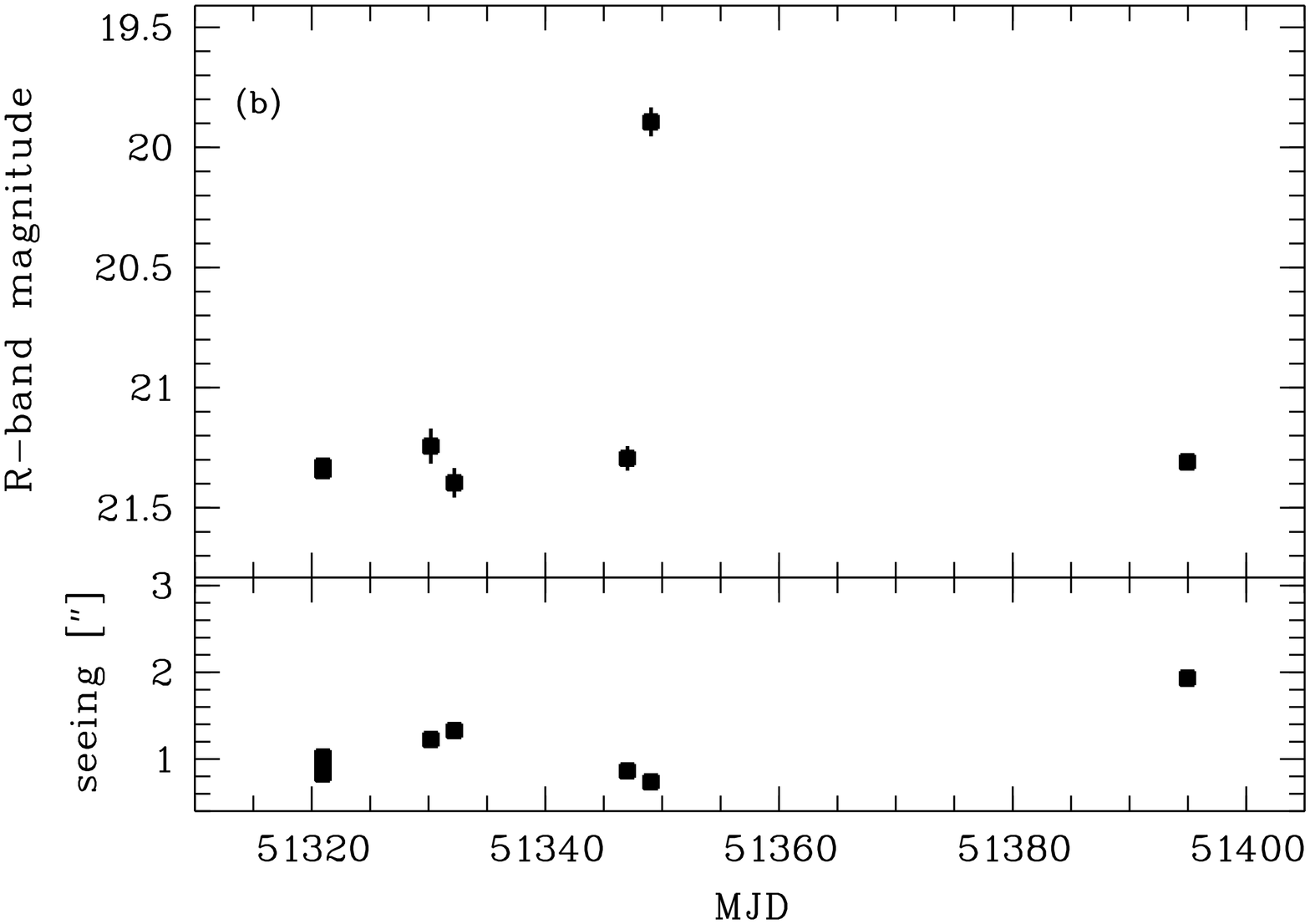}     
\hspace{-0.65cm}
\includegraphics[width=0.365\linewidth,angle=-90]{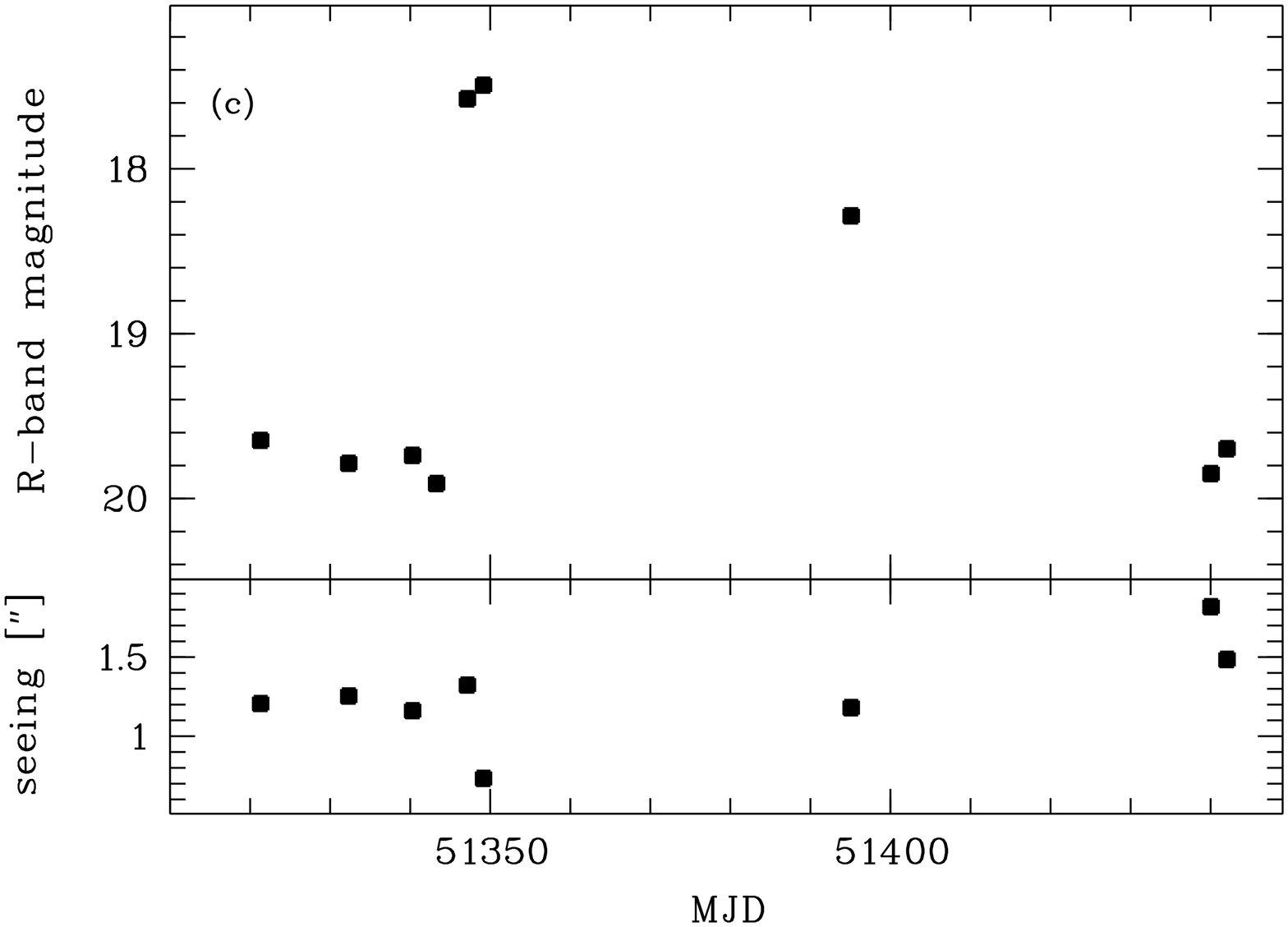}     
\hspace{-0.65cm}
\includegraphics[width=0.365\linewidth,angle=-90]{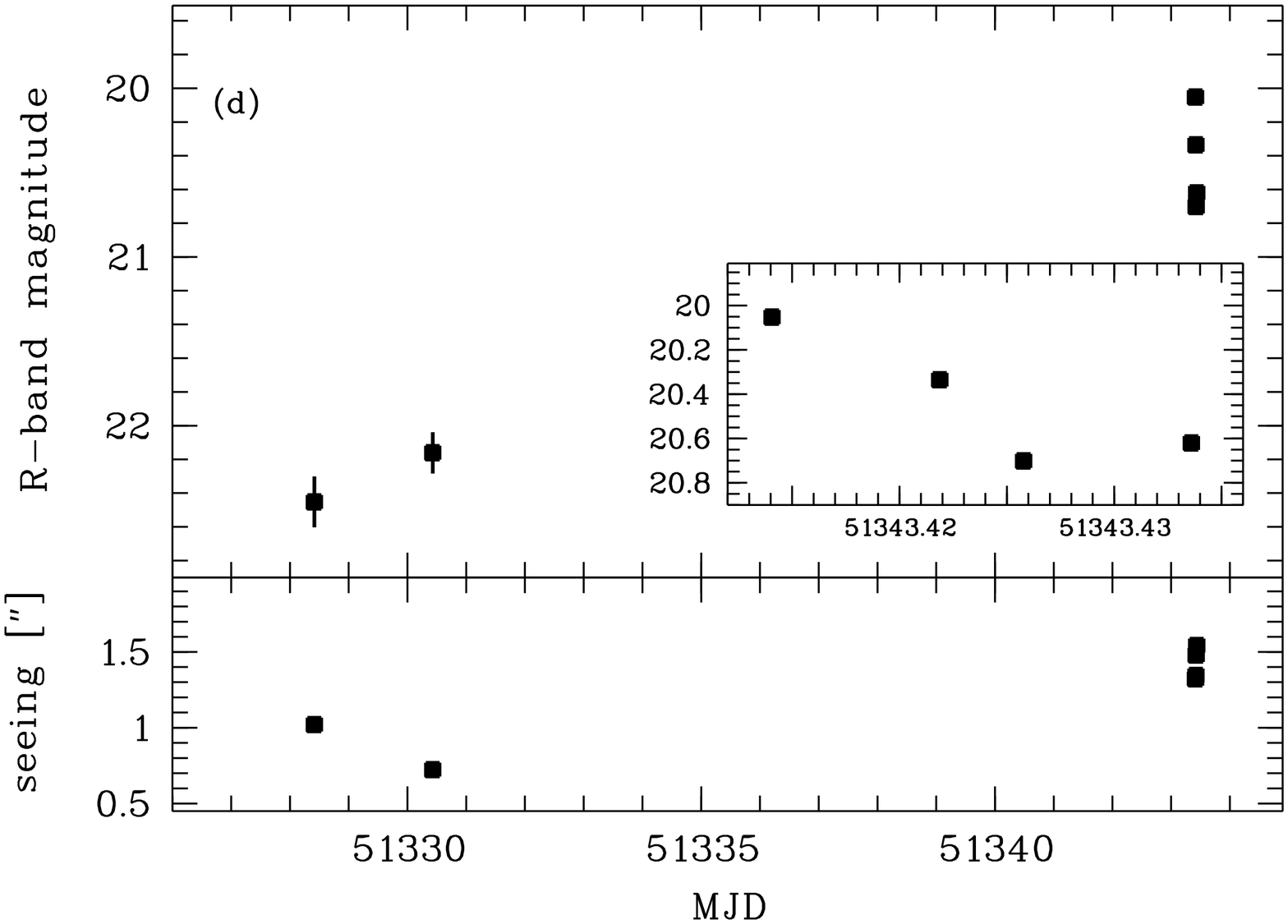}     
\caption{Light curves of the four transient source candidates (top) and seeing
evolution  (bottom).  Error  bars  are smaller  than  the symbol  size if  not
visible.  {\bf top left:} The candidate cataclysmic variable J132653.8-212702.
{\bf  top right:} J132813.7-214237,  extragalactic transient  candidate.  {\bf
bottom   left:}  J161953.3+031909,   a  dwarf   nova.   {\bf   bottom  right:}
J215406.6-274226,  a possible  flare star  or afterglow  candidate.  The inset
shows the detailed light curve at the time of the flare.}
\label{fig:cand}
\end{figure*}

The  source was initially  detected in  F3 during  the first  night of
observations on May 22, 1999 (MJD 51321.0045).  The field was observed
four times  during that night  and the source exhibited  a significant
brightness   increase  from  $R$=20.4   to  $R$=19.7   within  80\,min
(Fig.~\ref{fig:cand}a). It  was observed again on May  31/June 2, June
17/19 and August 4/6 and  showed a flaring of $\Delta R$=0.2--0.4\,mag
during  each of  these epochs.   No X-ray  or optical  counterpart was
detected in the ROSAT All-Sky  survey (taken 1990). No entry was found
in  SIMBAD or  NED.   The rapid  variability  excludes slowly  varying
objects and is consistent with a cataclysmic variable and/or eclipsing
binary. No orbital period could be obtained due to the sparse sampling
of the  light curve.   The confirmation of  the nature of  this source
requires  further  monitoring   of  the  variability  or  simultaenous
observations in multiple colours.
  
% --------------------------------------------------------------------------
% --------------------------------------------------------------------------

% --------------------------------------------------------------------------
\begin{figure*}
\begin{center}
\fbox{\includegraphics[width=0.455\linewidth,angle=0]{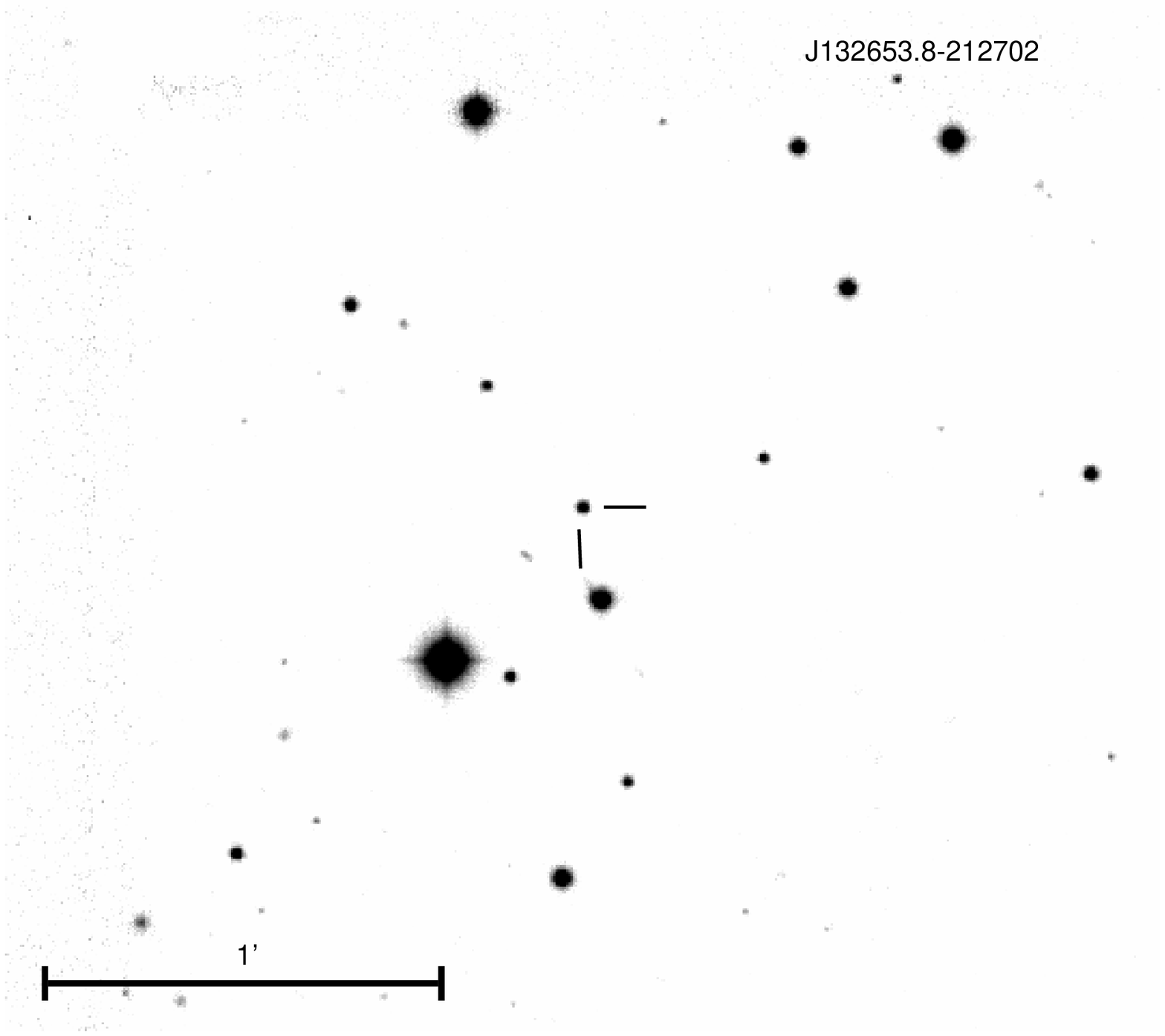}}
\hspace{0.1truecm}
\fbox{\includegraphics[width=0.455\linewidth,angle=0]{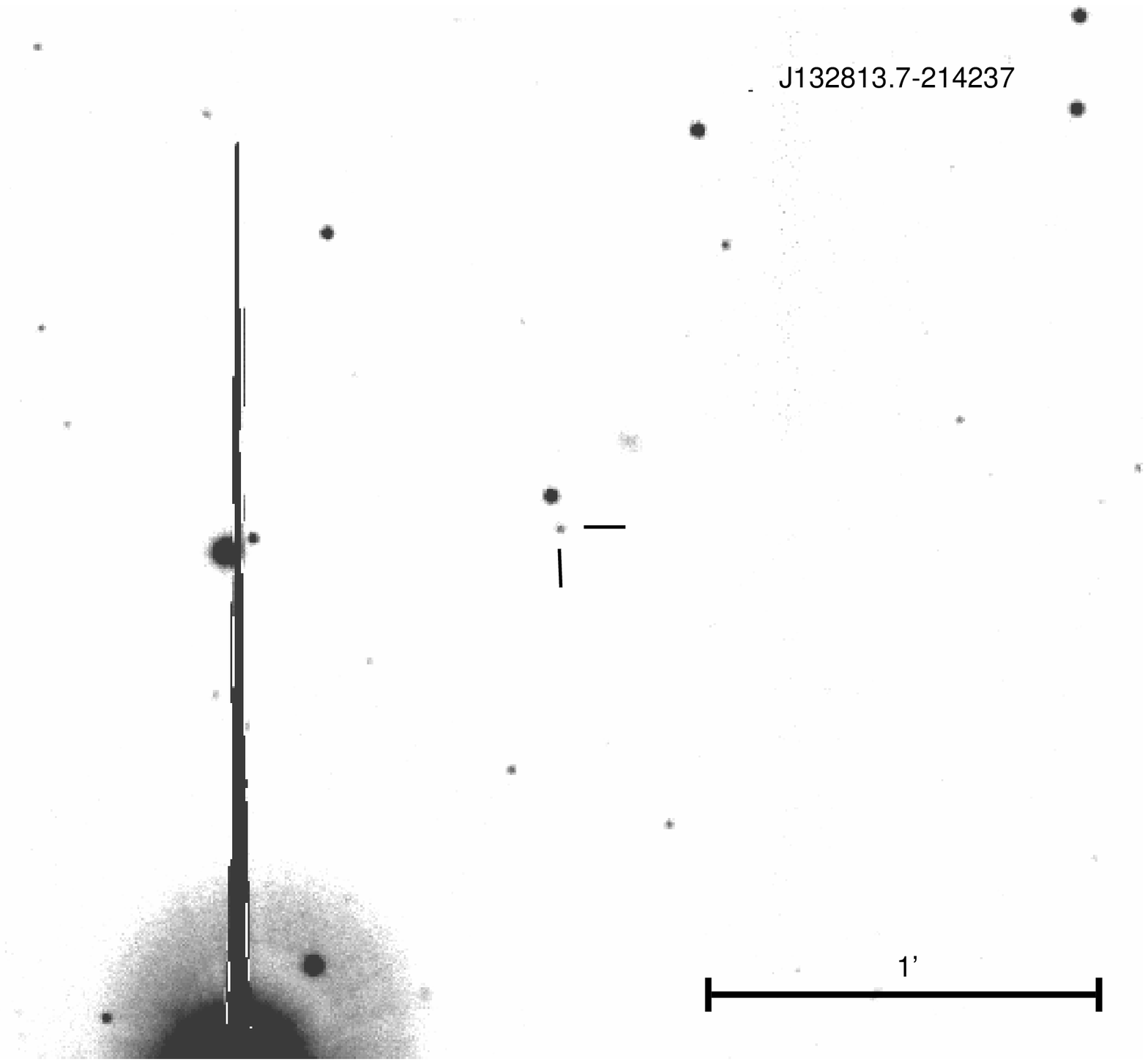}}     

\vspace{0.2truecm}

\fbox{\includegraphics[width=0.455\linewidth,angle=0]{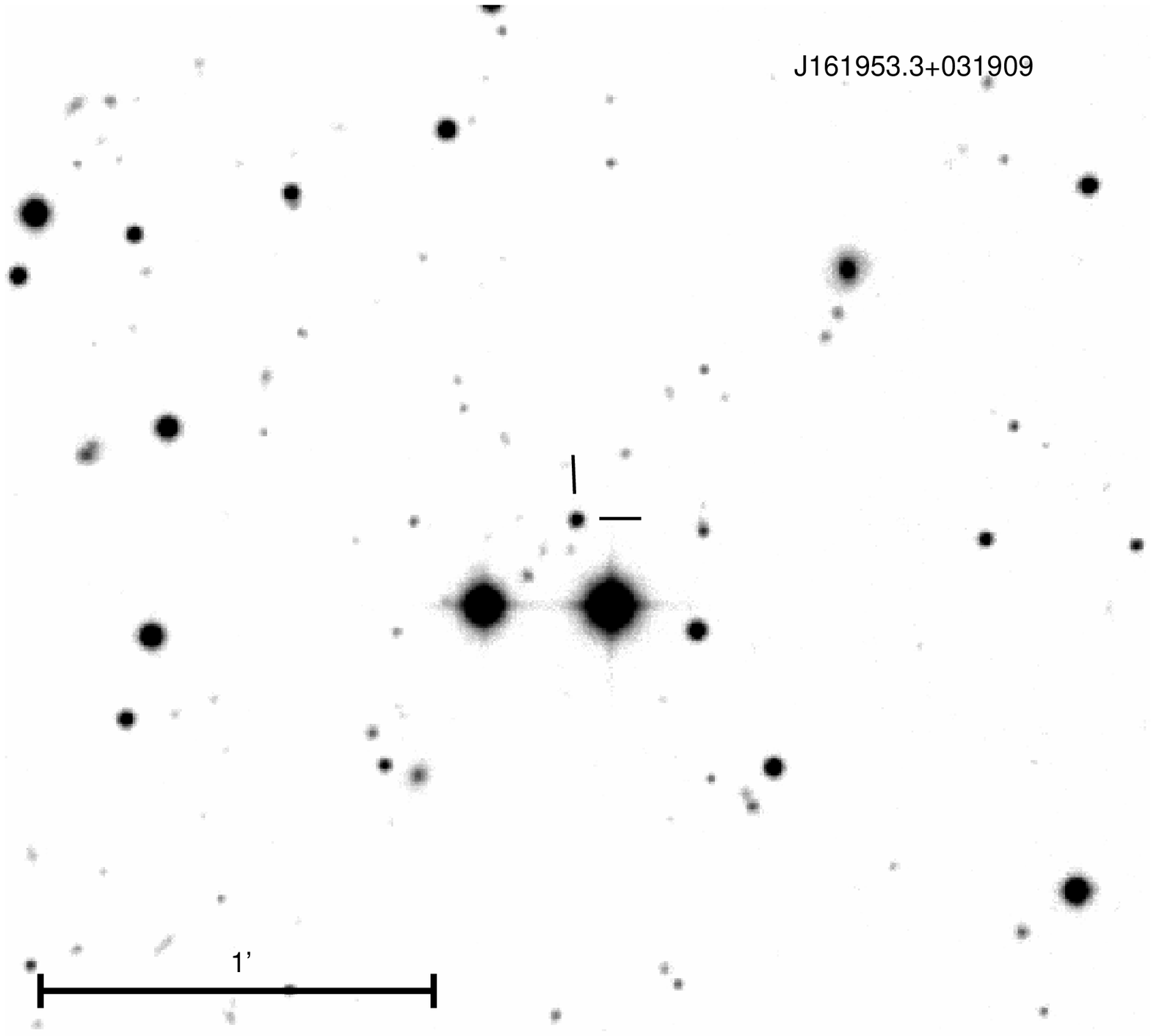}}     
\hspace{0.1truecm}
\fbox{\includegraphics[width=0.455\linewidth,angle=0]{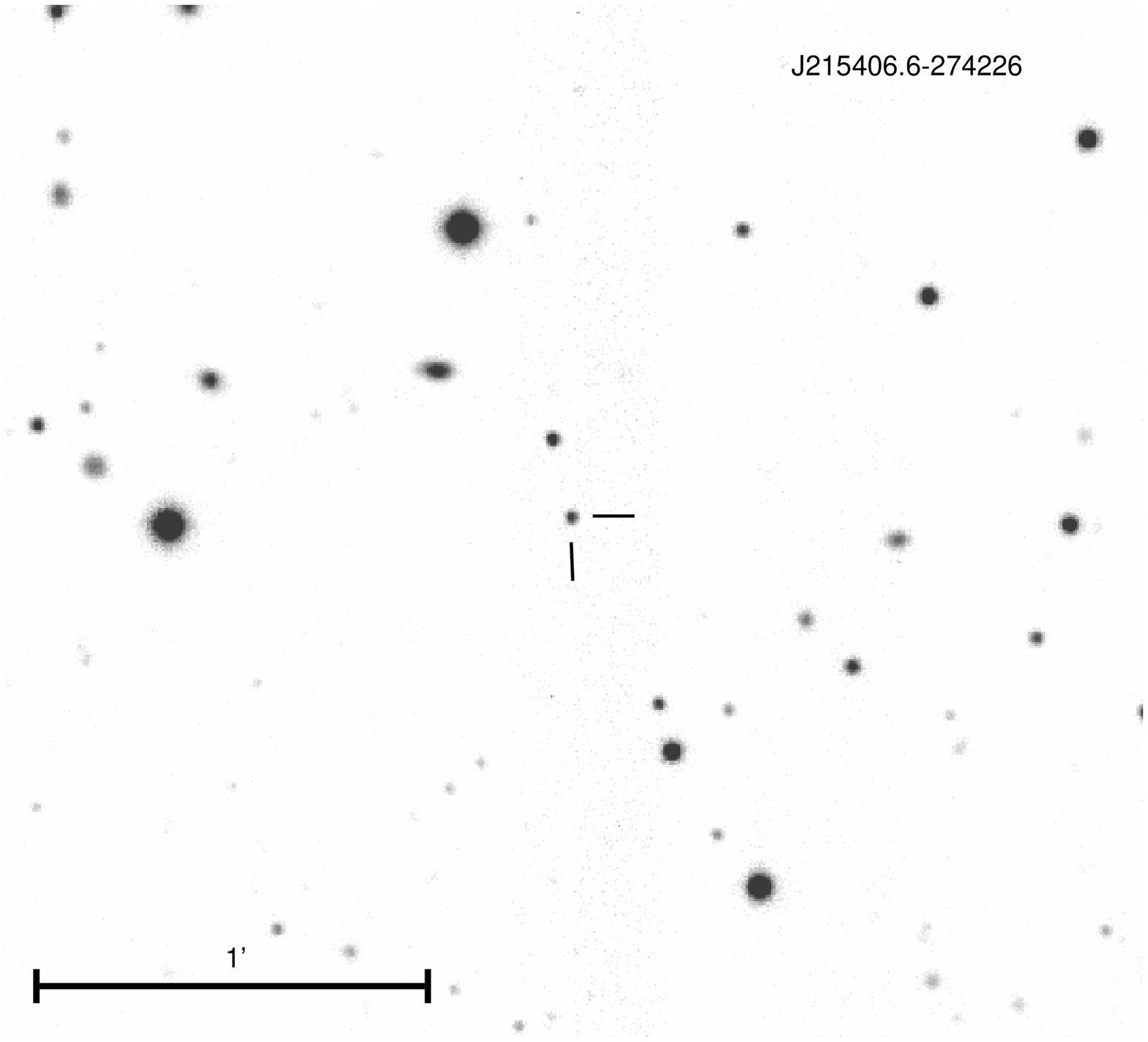}}     
\end{center}
\caption{Finding charts for  the four discovered transients. From  top left to
  bottom   right:    J132653.8-212702,   J132813.7-214237,   J161953.3+031909,
  J215406.6-274226. North is up and East to the left. }
\label{fig:candFc}
\end{figure*}
% --------------------------------------------------------------------------

\subsubsection{J132813.7-214237}

\begin{figure}[h]
\includegraphics[width=0.495\linewidth,angle=0]{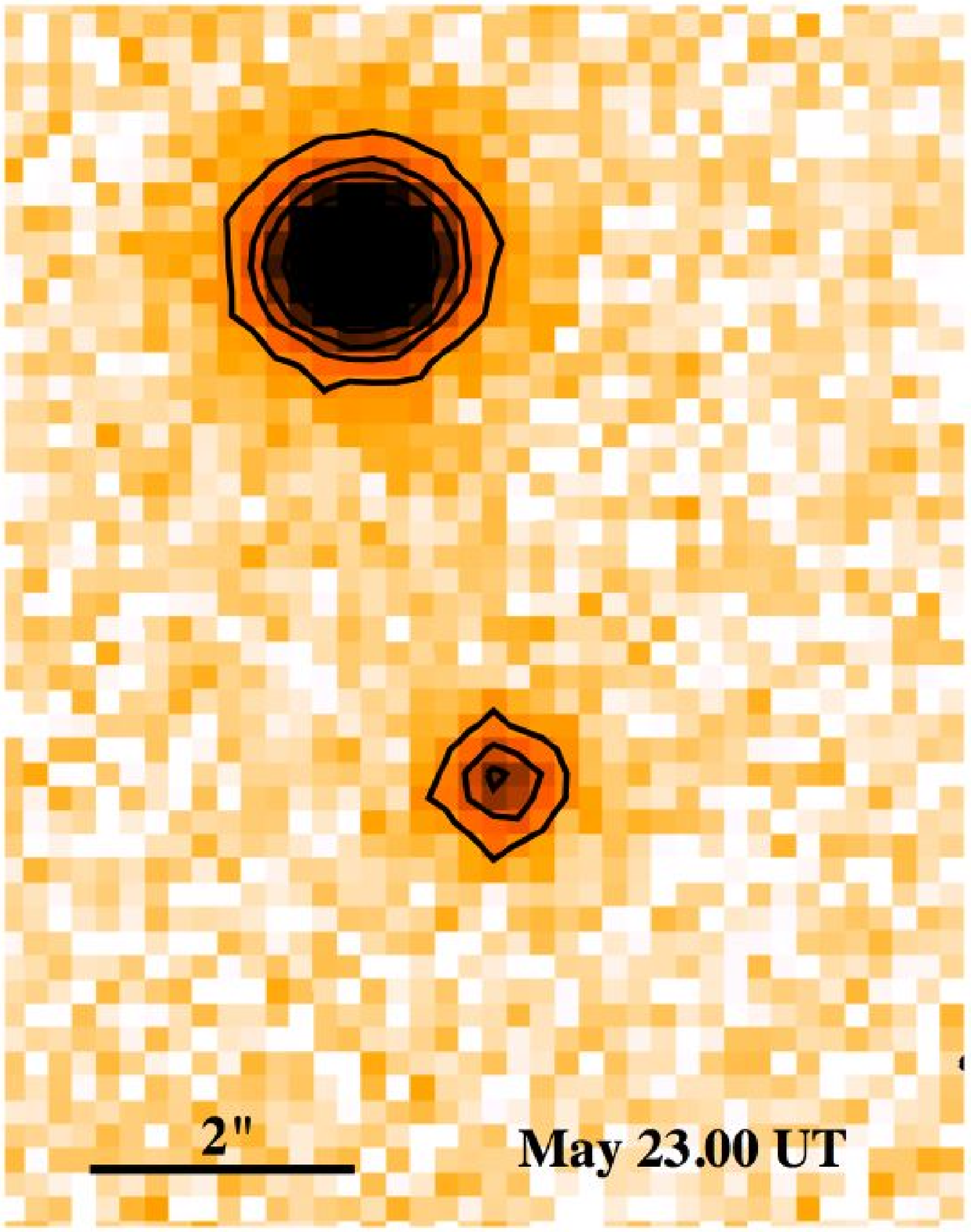}     
\includegraphics[width=0.495\linewidth,angle=0]{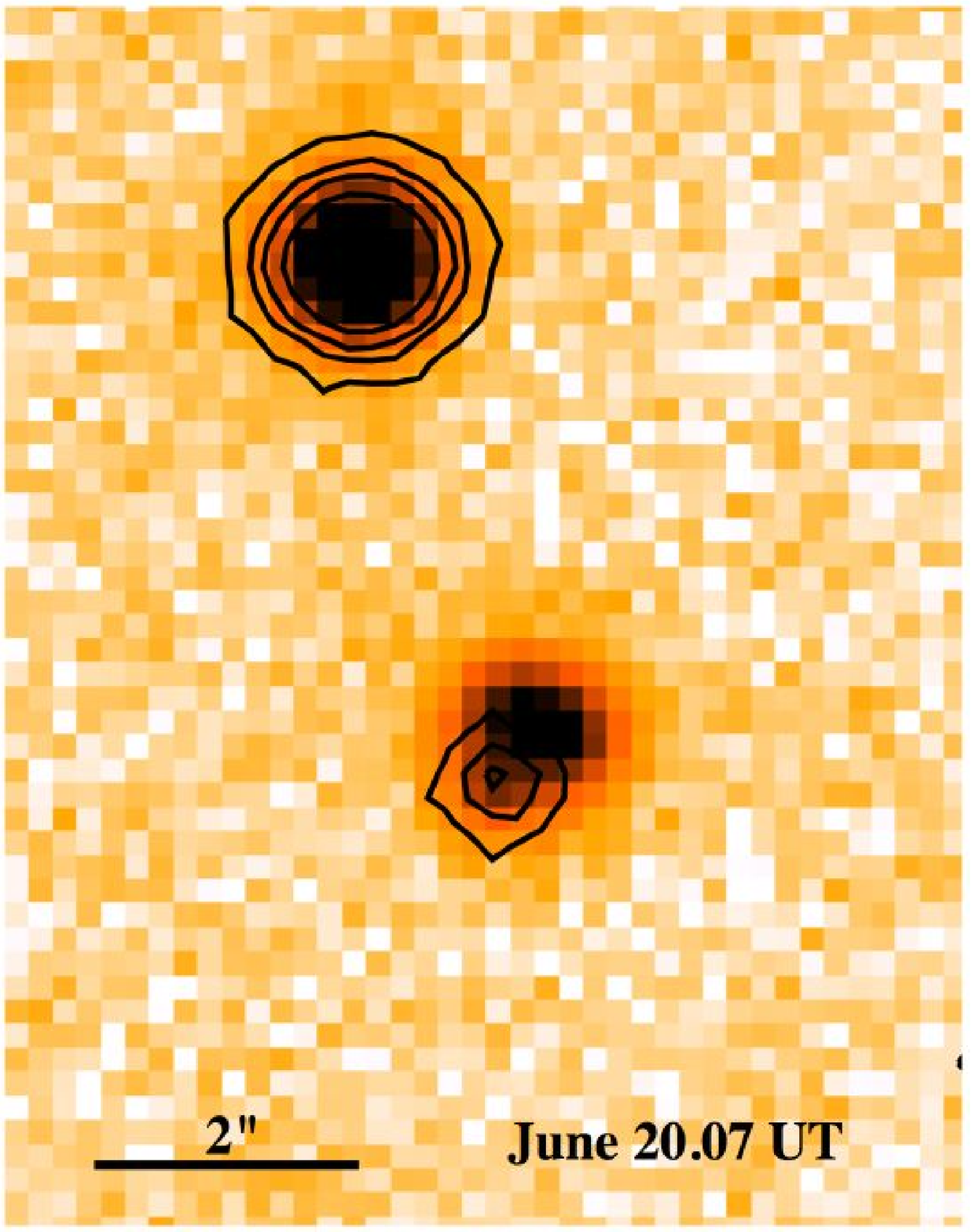}          
\caption{{\bf left:} Image of  J132813.7-214237 (center source) and a
  nearby bright star from an observation taken on May 23.00 UT when the source
  had  a  brightness  of  $R$=21.   Flux contours  are  overplotted  for  both
  objects. North is up and East  to the left. {\bf right:} Same field observed
  on June  20.07 UT  together with the  flux contours  from May 23.00  UT. The
  flaring  source  shows an  offset  of  $\sim$0\farcs8  with respect  to  the
  quiescent counterpart.}
\label{fig:kurz1-cont}
  \vspace{-0.5cm}
\end{figure}

The source was detected on June  20.07 UT 1999 (MJD 51349.0430) at its maximum
brightness  of $R$=19.9  (Fig.~\ref{fig:cand}b). Unfortunately,  the observing
period finished  after the observation of  the outburst and  the next $R$-band
pointing was performed  more than six weeks later on August  04. Thus, a decay
of  the source could  not be  monitored. A  faint ($R$=21.3)  persistent point
source was found  at the position of the transient in  images taken before and
after the  outburst. The  ROSAT All-sky survey  did not  show a source  at the
position   of   J132813.7-214237   at    3$\sigma$   flux   upper   limit   of
5.5$\times$10$^{-13}$\,erg cm$^{-2}$ s$^{-1}$.   Similarly, no counterpart was
found in the SIMBAD and NED databases. The object is visible near the limiting
magnitude of  the DSS  in the  $B$ and $R$-band  and not  detected in  the DSS
$I$-band.  A detailed  analysis revealed that the transient  source was offset
by   $\sim$0\farcs8  from   the   position  of   the  persistent   counterpart
(Fig.~\ref{fig:kurz1-cont}).   This might suggest  an extragalactic  origin of
the transient  assuming the  persistent counterpart can  be associated  with a
candidate  host galaxy.   Nevertheless,  this vague  assumption  based on  the
available observational  data requires a confirmation by  an accurate distance
measurement.

The   lacking   observational  coverage   of   the   decay   light  curve   of
J132813.7-214237 leaves the identification of the origin of the source an open
question. Assuming  the extraggalactic nature of the  faint persistent source,
the  flaring  source could,  for  instance,  be  associated with  a  supernova
explosion  occuring  in this  galaxy.  The  brightening  of $\sim$1.5\,mag  in
$\sim$2\,days is very steep compared to observed supernovae though (Leibundgut
\etal 1991).  Another explanation is that  of a foreground flare star close to
the  line of  sight  towards the  persistent  background source.  Furthermore,
J132813.7-214237 might be a potential orphan afterglow.  In order to test this
hypothesis  we searched  for triggered  GRBs  which occurred  during the  time
between the preceding observation (June 18.05 UT) and the outburst.  We found
5    cataloged\footnote{http://grbcat.gsfc.nasa.gov/grbcat/grbcat.html}   GRBs
during  this period  (BATSE  \#7609 \&  7610  and IPN  \#2066,  2067 \&  2069;
K.Hurley,  private communication).  None  of those  has a  position consistent
with J132813.7-214237 which allows to  exclude an association with a triggered
GRB.  Regardless, J132813.7-214237 can also be due to an untriggered or orphan
afterglow and is the best such candidate found during our survey.

% --------------------------------------------------------------------------
% --------------------------------------------------------------------------

\subsubsection{J161953.3+031909} 

This candidate was detected as a constant source ($R$=19.9) during the
first  four observations  of the  field F4  in May  and June  1999 and
exhibited  a sudden  brightening  by $\Delta  R$=2.4\,mag between  the
pointings  on  June 14.29  UT  (MJD 51343.2950  )  and  June 17.14  UT
(Fig.~\ref{fig:cand}c).   The following observation  on June  19.15 UT
showed  the  source  unchanged   and  later  observations  indicate  a
subsequent decay  over 50--90\,days back to  the quiescent brightness.
The   optical    light   curve   suggests    the   classification   of
J161953.3+031909  as  a  dwarf  nova.  This  is  strengthened  by  the
detection of a faint X-ray  source in the ROSAT all-sky survey. During
an exposure  of 320\,sec  on August  12/13, 1990 a  total of  9 source
photons  were detected, corresponding  to a  mean vignetting-corrected
count  rate of  0.033\,cts s$^{-1}$  (this is  below  the significance
threshold  of  the all-sky  survey  catalog,  so  this source  is  not
contained  in the  1RXS  catalog  of Voges  \etal  1999).  Adopting  a
thermal bremsstrahlung  model with 1\,keV temperature and  half of the
Galactic foreground absorption, an unabsorbed flux in the 0.1-2.4\,keV
band  of (7.0$\pm$1.0)$\times$10$^{-13}$\,erg  cm$^{-2}$  s$^{-1}$ (or
(9$\pm$2)$\times$10$^{-13}$\,erg  cm$^{-2}$  s$^{-1}$  bolometric)  is
derived.  Using the quiescent optical brightness, this implies a ratio
of L$_X$/L$_{opt}$ = 0.6, consistent  with SU UMa stars (Verbunt et al
1997).    With    an   X-ray   luminosity    of   1.1$\times$10$^{32}$
[D/1\,kpc]\,erg  s$^{-1}$, the  implied  distance is  of  order a  few
hundred parsec.  No entry was found in the SIMBAD and NED databases.

\subsubsection{J215406.6-274226} 

This source was  initially detected in two  epochs of the field F7  as a faint
object  with $R$=22.5  (Fig.~\ref{fig:cand}d).   On June  14.41  UT 1999  (MJD
51343.4150) it  was   found  $\Delta  R$=1.9\,mag   brighter  than  previously
measured.  Fortunately, during this night  four images of the field were taken
and  the rapid  fading of  J215406.6-274226 by  $\Delta R$=0.7\,mag  in around
20\,min,  corresponding to  a decay  with t$^{-1}$,  was discovered  (inset of
Fig.~\ref{fig:cand}d).  Unfortunately,   the  further  fading   could  not  be
monitored as the position of the source fell into the gap between two CCDs for
the following  twelve pointings.   While the rapid  decay would  be consistent
with the  observations of  early GRB  afterglows, also a  flare star  offers a
possible  explanation.  Given  the  range of  absolute  magnitudes for  nearby
M-dwarfs of  $M_V$=12--16 (Reid \etal 1995) the  observed quiescent brightness
would  place J215406.6-274226  at a  distance of  0.2--1.2\,kpc.  No  X-ray or
optical counterpart  was detected  in the ROSAT  All-Sky survey and  SIMBAD or
NED, respectively.  No  source is detected in the DSS  $B$ and $R$-band images
but  a faint source  ($\sim$7$\sigma$) is  visible in  the DSS  $I$-band.  The
quiescent counterpart  could not be resolved and  further observational effort
is required for a more solid classification of the object.

\section{Efficiency for Orphan Afterglow Detection}

The detection efficiency for on-axis  optical afterglows was estimated using a
set  of Monte-Carlo  simulations folded  with  the observing  schedule of  the
survey.   We simulated  afterglows with  random sky  coordinates,  light curve
parameters  and  explosion  times   distributed  over  the  periods  in  which
observations were taken.  For all  afterglows with positions inside one of the
monitored fields, the  expected magnitudes of the afterglows  in the first and
second observation of the field after the burst were calculated.

The  afterglow  light  curves  were  described  by  a  broken  power  law  and
parametrized by the pre-break slope, $\alpha_1$, break time, $t_b$, post-break
slope, $\alpha_2$ and initial $R$-band magnitude, $R_{in}$.  Here, the initial
magnitude corresponds to the observer  frame $R$-band brightness with which an
afterglow is created  in the simulations.  We used the  parameter ranges of 38
observed optical  afterglow light curves summarized  in Zeh et  al. (2005).  A
relatively flat  decay plus  a steeper  flux decrease after  the jet  break is
typically   observed   for  optical   afterglows   associated  with   detected
long-duration  GRBs  and is  also  expected for  on-axis  orphans  as well  as
``regular'' afterglows of untriggered  bursts.  As described earlier, off-axis
orphans will show a different behavior.   They are expected to be fainter than
on-axis  orphans at early  times and  follow the  post-break decay  of on-axis
afterglows  after an  initial phase  of  re-brightening.  It  appears safe  to
assume that on-axis afterglows will be the majority at a limiting magnitude of
$R$=19--21 and off-axis orphans will  dominate only at lower magnitudes (Nakar
\& Piran 2003).   As the survey strategy foresees to  consider only sources as
candidate orphans with at least  two detections spaced by two nights, off-axis
orphans with maximum brightness of $R$$>$21  will in most cases be to faint to
be identified.   While bright  off-axis orphans could  in principle  have been
detected in  the survey, we do not  include them in the  simulations and focus
solely on untriggered and on-axis orphans.

The  expected number  of  detected afterglows  in  the survey  depends on  the
choices for  $\alpha_1$, $\alpha_2$, $t_b$  and $R_{in}$.  The ranges  for the
slopes  and break time  were obtained  from observed  light curves  (Zeh \etal
2005).      Accordingly,     we     used    Gaussian     distributions     for
--1.8$<$$\alpha_1$$<$--0.4  and   --2.8$<$$\alpha_2$$<$--1.4  and  an  uniform
distribution  for  $t_b$  ranging  from 0.4--4\,days.   The  most  influencing
parameter for  the outcome of  the simulations is  the initial magnitude  of a
candidate afterglow. Rapid optical follow-up observations of BATSE bursts with
LOTIS and ROTSE-I (Park \etal 1999;  Akerlof \etal 2000; Kehoe \etal 2001) and
recent {\it  Swift}/UVOT detections indicated that the  preponderance of early
afterglows does not get  brighter than $R$$\sim$14.  Furthermore, {\it HETE-2}
follow-up and  UVOT observations showed  that around 50\,\% of  the afterglows
might be  brighter than $R$$\sim$18.5  for 30\,minutes after the  burst (e.g.,
Lamb \etal 2004).

In  order to test  the influence  of the  limits of  $R_{in}$ on  the expected
number of afterglows  in the survey, we performed  two simulations with 10$^6$
bursts per  year and full sky,  each.  The initial  magnitudes were uniformly
distributed  between 9$<$$R_{in}$$<$20  and  13$<$$R_{in}$$<$23, respectively.
In addition,  a distribution proportional to 0.2$\times$$R_{in}$  in the range
of  13$<$$R_{in}$$<$23   was  simulated.    The  latter  corresponds   to  the
observations of  afterglows presented in Zeh  \etal (2005) and  is expected to
reproduce  the  reality  more  closely  than an  uniform  distribution.   Each
simulation was repeated  10$^3$ times and the mean detection  rates at a given
$R$-band magnitude were obtained.

\begin{figure}[h]
\includegraphics[width=0.73\linewidth,angle=-90]{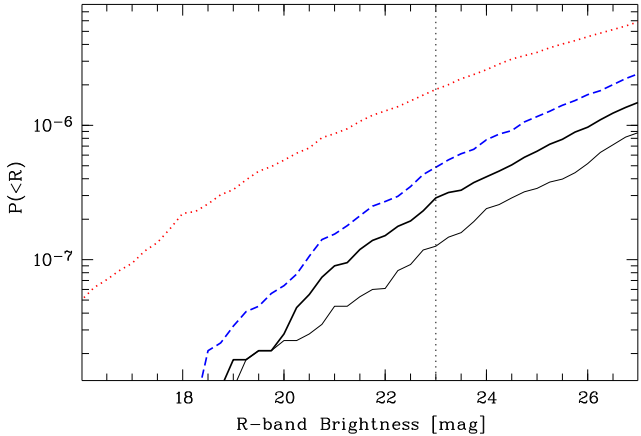}     
\caption{Probability  distributions  for  the detection  of  a  randomly
  selected afterglow  for three different  models of the  initial brightness.
  The  dotted   and  dashed  functions  represent   uniform  distributions  of
  $R_{in}$=9--20 and  $R_{in}$=13--23, respectively. The thick  and thin solid
  lines  correspond  to the  first-time  and  second-time  observations for  a
  distribution   proportional   to  0.2$\times$$R_{in}$   in   the  range   of
  $R_{in}$=13--23. The  vertical dotted line  marks the limiting  magnitude of
  the presented survey.}
\label{fig:magDis}
  \vspace{-0.5cm}
\end{figure}

The results of the simulations are presented in Figure~\ref{fig:magDis}.  Here
we  show the  normalized (devided  by the  total number  of  simulated events)
probability for a first-time detection  of a simulated afterglow brighter than
a given magnitude.  The normalized probability represents the chance to detect
the afterglow  in a simulation  with one single  event over the year  and full
sky. The expectation for a second observation of a candidate afterglow for the
model  with  non-uniform  distribution   of  $R_{in}$  is  included  as  well.
Naturally, the probability to detect  an afterglow increases with the depth of
the  survey. More  transients  are expected  to  be detected  above a  certain
magnitude for afterglow models with brighter $R_{in}$. Due to the lower number
of bright  afterglows and the increasing probability  with diminishing initial
brightness, the  non-uniform distribution gives the  lowest expectation rates.
At  the limiting magnitude  of the  survey of  $R$$\sim$23 the  probability to
detect a specific  randomly selected afterglow in at  least one observation is
approximately 3$\times$10$^{-7}$.   The detection rate  of events in  at least
two consecutive  observations is  lower by  a factor of  $\sim$3. In  the more
optimistic  models, the  probability  for  a single  detection  reaches up  to
2$\times$10$^{-6}$.

% --------------------------------------------------------------------------
% --------------------------------------------------------------------------

\section{Discussion}

The Monte Carlo simulations described  in the previous section provide us with
the number of  afterglows (per year and full  sky), $N_{MC}$, which correspond
to the probability of identifying one event in two consecutive observations of
the survey.  For  the three tested afterglow parameterizations  we find values
between $N_{MC}$=1.5$\times$10$^{6}$ and $N_{MC}$=1$\times$10$^{7}$.

Throughout the  survey, four  unidentified optical transients  were discovered
and one  of these sources shows  indications for an  extragalactic origin (see
Sect.~3.1.2).  Nevertheless, the flaring of  the source was only detected in a
single observation and thus the fading could not be monitored.  An unambiguous
identification of the transient with a GRB afterglow is therefore not possible
which  leaves us  with the  result of  having no  clear orphan  or untriggered
afterglow  detected  in  the  data.   Therefore, $N_{MC}$  obtained  from  the
simulations can be interpreted as an upper limit on the true number of on-axis
afterglows  per year and  full sky.   For $N$$>$$N_{MC}$  one or  more on-axis
orphans would have been expected in the data.

As  advertised earlier,  on-axis  orphan afterglows  can  be used  to place  a
constraint  on  the  collimation  of  the optical  afterglow  emitting  region
relative to the collimation of the $\gamma$-ray emitting jet.  The collimation
factor, $f_c$, corresponds  to the ratio of the true  rates of on-axis optical
afterglows, $N_{A}$,  and long-duration GRBs which  produce observable optical
afterglows,  $N_{\gamma}$, pointed  at the  Earth.  With  $N_{MC}$$>$$N_A$, an
upper    limit     for    the     collimation    can    be     derived    from
$f_c$$<$$N_{MC}$/$N_{\gamma}$.   Obtaining  $N_{\gamma}$  is  not  trivial  as
several uncertain  factors influence the  number of GRBs for  which afterglows
are  in principle  detectable.   In  a simplified  model  $N_{\gamma}$ can  be
written as

\begin{equation}
N_\gamma=N_{\gamma,obs}\cdot f_X \cdot f_D \cdot f_S
\end{equation}

where $N_{\gamma,obs}$ is the observed rate of long-duration GRBs with
a  specific  $\gamma$-ray   instrument  corrected  for  sky  coverage,
Earth-blockage  and  instrument down-time  $f_X$  corrects for  events
outside the instrument's energy  range, $f_D$ is the correction factor
for optically  dim or dark  afterglows of long-duration  bursts (e.g.,
intrinsically  faint,  absorbed,  high-$z$)  and  $f_S$  corrects  for
possible afterglows associated with  short bursts.  Using the full sky
GRB rate  measured with BATSE  of $\sim$666 yr$^{-1}$  (Paciesas \etal
1999) and  correcting for  the ratio  of  long to  short bursts  (2:1;
Kouveliotou  \etal 1993),  we  obtain $N_{\gamma,obs}$$\sim$444.   The
{\it HETE-2}  population of high energy bursts  revealed a composition
of X-ray flashes (Heise \etal  2001), X-ray rich bursts and ``normal''
GRBs in equal  parts (Lamb \etal 2005).  X-ray  flashes and X-ray rich
bursts show  similar afterglows as  observed for GRBs but  have softer
prompt  emission  spectra.   As  BATSE  was less  sensitive  at  lower
energies than {\it HETE-2} a  fraction of these events might have been
missed.  This could require a correction factor as large as $f_X$=2.

In general, not for all  rapidly followed GRBs an associated optical transient
can be found.  Some detected afterglows were faint already  early on and would
have fallen below  the limiting magnitude reached in  our survey. The fraction
of these events was found to be of the order of 10\,\% (e.g., Lamb \etal 2004;
Jakobsson  \etal 2004; Rol  \etal 2005)  which provides  $f_D$$\sim$0.9.  Only
recently the first optical transients of short-duration GRBs were found (Price
\etal  2005; Gal-Yam  \etal 2005).   Theory and  observations hint  that these
afterglows are  significantly fainter  than the counterparts  of long-duration
bursts.  Therefore,  we neglect  the influence of  short burst  afterglows and
apply $f_S$=1 (as we also subtracted them from the burst rate of BATSE).

Using    the    assumptions   discussed    above    together   with    $N_{MC}
$=1$\times$10$^{7}$   we    derive   $N_\gamma$$\sim$800   and   $f_C$$<$12500
accordingly.   This rather  conservative upper  limit is  significantly higher
than the  beaming correction  derived by Guetta  \etal (2005) and  Frail \etal
(2001) of  75$\pm$25 and  500, respectively.  The $\gamma$-ray  beaming factor
corresponds to  the ratio of  the overall rate  of GRBs to the  detected burst
rate and  should be  an upper limit  for $f_C$  (corresponding to the  case of
isotropic afterglow radiation).

The  high upper  limit  on $f_C$  shows  that the  effective  coverage of  the
performed observations were not sufficient  to provide a strong constraint for
the  collimation.   An  approximately  25  times larger  portion  of  the  sky
($\sim$325\,deg$^2$) would  have been  required to reduce  $f_c$ to  less than
500,  assuming no  orphan afterglow  detection. We  performed a  further Monte
Carlo simulation in  order to estimate the properties  for an ``ideal'' survey
assuming a  schedule of one  observation per field  every two nights  over 150
nights with a  limiting magnitude of $R$=23. We  find that $f_C$$<$500 ($<$75,
$<$10)  would  be  reached  with   such  a  configuration  and  a  50\,deg$^2$
(300\,deg$^2$, 2500\,deg$^2$)  field. Although ambituous, a  program like this
is  in  the  range  of  the  near-future  instrumentation  (e.g.,  VLT  Survey
Telescope,  Visible   \&  Infrared  Survey  Telescope   for  Astronomy)  which
encourages to perform a comprehensive  search for untriggered GRBs in the near
future.   In  addition,  spectroscopic  follow-up  observations  of  candidate
orphans would be important to distinguish between afterglows and other optical
transients.

The non-detection of  an afterglow also provides a limit on  the rate of other
explosive  events with  similar fading  behaviour. This  includes  events with
minor or quashed  high energy emission like failed GRBs  (Huang \etal 2002) or
so-called ``dirty fireballs'' (Dermer \etal 1999).

\section{Conclusion}

We presented  the data reduction, analysis  and results of  an $R$-band survey
dedicated  to slowly  variable optical  transients.  The  survey  strategy was
designed  specifically  to  search  for afterglows  of  untriggered  gamma-ray
bursts. 12\,deg$^{2}$  were monitored in up  to 25\,nights down  to a limiting
magnitude  of   $R$=23.   Throughout  the  survey,   four  previously  unknown
transients  were  discovered.  Based  on  the limited  photometric  data  only
putative  classifications of  the candidates  could be  obtained so  far.  The
observations  of three  of the  transients suggest  them to  be  a cataclysmic
variable, a flare star and a dwarf nova, respectively.

The fourth  transient appeared in a  single image as a  bright source slightly
offset with respect to an underlying  quiesent object. The decay of the source
could not be followed  due to the lack of observations, thus  the orgin of the
transient is unresolved. However,  the spatial association with the persistent
source  suggests an  extragalatic  origin. The  steep  brightness increase  of
$\sim$1.5\,mag  in $\sim$2\,days appears  a-typical for  a supernova  and thus
makes the detected  flaring source the best candidate  for an orphan afterglow
in our survey.

Simulations of the transient detection efficiency of the survey showed
that the effective sky coverage  was not sufficient to obtain a strong
constraint on  the collimation of  the optically emitting  GRB outflow
from  the  non-detection of  suitable  counterparts. Nevertheless,  we
found that a similar programs like the one described in this paper are
feasible to be performed in the near future. Limits on the collimation
ratio of the X-ray to gamma-ray emitting regions were already obtained
in the  past and strong ratios  ($>$8) were ruled  out (Grindlay 1999;
Greiner  \etal 2000).   Tighter  constraints on  the optical  emitting
region  will become  available  soon as  well  with the  use of  large
dedicated surveys.

% --------------------------------------------------------------------------
\bigskip

\noindent
{\it Acknowledgments.}  We  thank the many observers who  have assisted in the
completion of this survey, including M.  Braun, D.  Clowe, J.  Eisl\"offel, J.
Fried, P.   Heraudeau, R.   Klessen, T.  Kranz,  I.  Lehmann, R.   Schmidt, C.
Wolf, D.  Woods  and the helpful night assistants at La  Silla. We thank Mario
Radovich (INAF - Osservatorio Astronomico di Capodimonte) for making the WIFIX
image  reduction  and astrometry  package  publicly  available.  We thank  the
anonymous referee for useful comments. The  WFI is a joint project between the
European  Southern Observatory,  the Max-Planck-Institut  f\"ur  Astronomie in
Heidelberg (Germany) and the Osservatorio Astronomico di Capodimonte in Naples
(Italy).

\bigskip

% --------------------------------------------------------------------------

\end{document}